%% file: main.tex
\documentclass{article}
\usepackage{spconf,amsmath,graphicx}
\usepackage{latexsym}
\usepackage{comment}
\usepackage{amssymb}
\usepackage{dsfont}
\usepackage{bm}
\usepackage{color}

\usepackage{multirow}
\usepackage{tabularx}
\usepackage{threeparttable}
\usepackage[export]{adjustbox}
\usepackage{multicol}
\usepackage{cite}
\usepackage{booktabs}
\usepackage{hhline}
\usepackage[format=hang]{caption}
\usepackage{algorithm}
\usepackage[noend]{algpseudocode}
\usepackage{arydshln}

\def\Underline{\setbox0\hbox\bgroup\let\\\endUnderline}
\def\endUnderline{\vphantom{y}\egroup\smash{\underline{\box0}}\\}

\title{Conversation-oriented ASR with multi-look-ahead CBS architecture}
\name{Huaibo Zhao, Shinya Fujie, Tetsuji Ogawa, Jin Sakuma, Yusuke Kida*, Tetsunori Kobayashi}
\address{Department of Communications and Computer Engineering, Waseda University, Tokyo Japan \\
*Line corporation, Tokyo Japan}

\begin{document}
%
\maketitle
\input{subtex/abstract.tex}

\input{subtex/intro}

\input{subtex/tech}

\input{subtex/proposal}

\input{subtex/exp}

\input{subtex/conclusion}




\fontsize{8.7pt}{9.6pt}\selectfont
\bibliographystyle{IEEEtran}
\bibliography{strings,refs}

\end{document}

%% file: subtex/abstract.tex
\begin{abstract}
During conversations, 
humans are capable of inferring the intention of the speaker at any point of the speech to prepare the following action promptly.
Such ability is also the key for conversational systems to achieve rhythmic and natural conversation.
To perform this, the automatic speech recognition (ASR) used for transcribing the speech in real-time must achieve high accuracy without delay.
In streaming ASR, high accuracy is assured by attending to look-ahead frames, which leads to delay increments.
To tackle this trade-off issue, 
we propose a multiple latency streaming ASR to achieve high accuracy with zero look-ahead.
The proposed system contains two encoders that operate in parallel, 
where a primary encoder generates accurate outputs utilizing look-ahead frames, and the auxiliary encoder recognizes the look-ahead portion of the primary encoder without look-ahead.
The proposed system is constructed based on contextual block streaming (CBS) architecture, which leverages block processing and has a high affinity for the multiple latency architecture.
Various methods are also studied for architecting the system, 
including shifting the network to perform as different encoders;
as well as
generating both encoders' outputs in one encoding pass.

\end{abstract}

%% file: subtex/intro.tex
\section{Introduction}
\label{sec:intro}

We humans can infer the intention of a speaker in conversation, 
even in the middle of an utterance, and prepare the following action to be taken. 
This function allows us to respond at the appropriate timing, 
sometimes without waiting for the end of an utterance, 
and to achieve a rhythmic and natural conversation.
To achieve this function in conversational systems,
the speech recognizer of the system is required to transcribe the input speech accurately without delay as any instant in time.
High accuracy can be achieved in speech recognition by applying look-ahead,
which provides the speech recognizer with more forward clues to make reliable decisions,
but leads to look-ahead latency,
which largely increases the delay of speech recognition.
This research aims to develop a highly accurate speech recognizer that can operate with zero look-ahead.

One recent trend in speech recognition develops around the end-to-end models~\cite{Graves2014TowardsES,Chorowski2015AttentionBasedMF,Chan2016ListenAA,watanabe2017hybrid}.
Among them, the recent Transformer-based methods achieve high performance by taking advantage of the self-attention function, but also require look-ahead in their structure~\cite{vaswani2017attention,Gulati2020ConformerCT,Dong2018SpeechTransformerAN,Karita2019ACS,Lscher2019RWTHAS,higuchi2021improved}.
Conversational speech recognition requires streaming ASR, but the look-ahead requirement also exists to guarantee the performance~\cite{Shannon2017ImprovedED,Chang2019JointEA,Chang2022TurnTakingPF}.
Although it can be implemented in causal by applying attention-mask to the look-ahead part~\cite{Povey2018ATS,sukhbaatar2019adaptive,Chang2020EndtoEndAW}, the degradation from full context implementations that allow look-ahead is significantly large~\cite{moritz2020streaming}.

In this vein, there are attempts in the multi-latency approach, which combines a short look-ahead ASR and a long look-ahead one~\cite{Mahadeokar2022StreamingPT,Narayanan2021CascadedEF,Sainath2021AnES,Shi2021DynamicET}.
In~\cite{Mahadeokar2022StreamingPT}, a high-latency encoder (long look-ahead) operates on the outputs of a low-latency (short look-ahead) encoder to correct the beam search results.
In~\cite{Sainath2021AnES}, a second-pass non-streaming recognition is conducted to refine the first-pass streaming outputs.
A common feature of them is the cascaded configuration, in which the high-latency, high-precision recognizers operate on the results of the low-latency recognizers to compensate for them.

Similarly, the proposed system in this study is a multi-latency ASR that combines a high-latency/high-accuracy encoder and a low-latency one. 
However, it is unique in that the system operates both encoders in parallel and adopts contextual block streaming ASR~\cite{Tsunoo2019TransformerAW, Tsunoo2021StreamingTA} (referred to as CBS), which has a high affinity with multi-latency architecture, as the base system.
In the proposed method, the decoder operates primarily with the output of the primary encoder, which works with the look-ahead. 
However, the look-ahead portion of the primary encoder, where there is no output from the primary encoder, is taken by the auxiliary encoder, which operates with zero look-ahead. 
Thus, the whole system constitutes a recognizer that operates accurately with zero look-ahead.

The CBS proposed by Tsunoo et al., the base system of our encoder, leverages block processing to achieve streaming properties in the attention-based encoder-decoder model architecture.
CBS contains a contextual block streaming encoder~\cite{Tsunoo2019TransformerAW}, which gracefully utilizes contextual information from the previous block and achieves high recognition accuracy.
Compared to frame-wise computation, block-wise processing in CBS possesses higher efficiency for using both primary and auxiliary encoders simultaneously.
With an easily adjustable look-ahead range in the block setting, CBS is also suitable for realizing parameter sharing of multiple encoders.

Since the original CBS (referred to as CBS-E/D) is based on the encoder-decoder architecture, the high computational cost in the decoder limits the real-time performance of CBS.
Hence, in this paper,
we also propose Transducer-based CBS (referred to as CBS-T), which combines the encoder of CBS and Transducer~\cite{Graves2012SequenceTW,Zhang2020TransformerTA,Chen2021DevelopingRS} to speed up the whole process, and also try to construct multi-latency ASR based on it.

The rest of the paper is organized as follows.
Section~\ref{sec:techniques} introduces our base models: CBS-E/D and CBS-Transducer.
In Section~\ref{sec:proposal}, we describe the proposed multiple latency streaming ASR system and provide various methods for constructing the system.
In Section~\ref{sec:exp}, we examine the effectiveness of the proposed method through speech recognition experiments and analyze the results.
Finally, Section~\ref{sec:conclusion} concludes this paper.

%% file: subtex/tech.tex
\section{Background}
\label{sec:techniques}
In this study, we adopt both contextual block streaming encoder-decoder (CBS-E/D) and contextual block streaming Transducer (CBS-T) as our base models.

\subsection{Contextual block streaming encoder-decoder}
\label{ssec:cbs}
\begin{figure}[tb]
\centering
\centerline{\includegraphics[width=\linewidth]{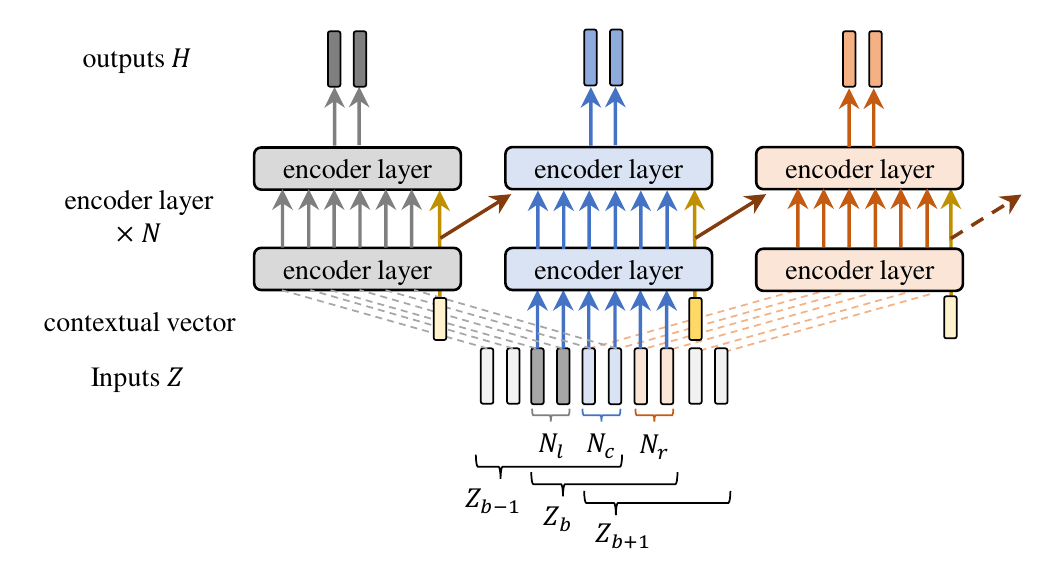}}
\caption{Block processing in CBS encoder}
\label{fig:cbs}
\end{figure}

As an attention-based encoder-decoder model, 
CBS-E/D conducts streaming processing in both encoding and decoding.
As shown in Fig.~\ref{fig:cbs}, 
for streaming encoding, CBS-E/D utilizes block processing with a context inheritance mechanism~\cite{Tsunoo2019TransformerAW}.
The speech input is segmented into blocks containing history, target, and look-ahead frames with the numbers of $N_l$, $N_c$, and $N_r$.
When a block is passed on to the encoder, the target frames are processed for the output with future contexts provided by the look-ahead frames, as well as history contexts provided by history frames and a contextual embedding vector inherited from the previous block.
Streaming decoding is achieved by a block boundary detection (BBD) algorithm~\cite{Tsunoo2021StreamingTA},
which examines the outputs' index boundaries and enables the beam search synchronous to the encoded blocks.
The streaming processing in CBS-E/D
is calculated as follows:
\begin{align}
&H_{b}, \bm{\mathrm{c}}_{b} = \mathrm{BlockEncoder}(Z_{b},\bm{\mathrm{c}}_{b-1}),
\label{eq:5}
\\
&\alpha(y_{0:i},H_{1:B})\approx\sum_{b=1}^{B} \sum_{j=I_{b-1}+1}^{I_b} \log p(y_i| y_{0:j-1},H_{1:b}).
\label{eq:6}
\end{align}
Eq.~\eqref{eq:5} represents the streaming encoding where the $b$-th input block $Z_{b}$ with $|Z_{b}| = N_l+N_c+N_r$ and a contextual vector from the previous block $\bm{\mathrm{c}}_{b-1}$ are processed to output the acoustic feature $H_{b}$ and current contextual vector $\bm{\mathrm{c}}_{b}$.
Eq.~\eqref{eq:6} represents the score of the partial hypothesis $y_{0:i}$ during streaming beam search decoding, where $y_{0}$ is the start-of-sequence token.
$I_b$ denotes the index boundary of the $b$-th input block derived from the BBD algorithm.

\subsection{Contextual block streaming Transducer}
\label{ssec:tt}
A CBS-T model combines the CBS encoder and the Transducer framework.
A Transducer framework contains three components: acoustic encoder, label encoder, and joint network.
Given a streaming input to a current time index $t$,
the output probability of each token is calculated as follows:
\begin{align}
&\bm{\mathrm{h}}_{t}^{\mathsf{AE}} = \mathrm{AcousticEncoder}(\bm{\mathrm{x}}_{1:t})\label{eq:1},
\\
&\bm{\mathrm{h}}_{u-1}^{\mathsf{LE}} = \mathrm{LabelEncoder}(y_{1:u-1})\label{eq:2},
\\
&\bm{\mathrm{h}} = \mathrm{Tanh}(\mathrm{Linear}(\bm{\mathrm{h}}_{t}^{\mathsf{AE}})+\mathrm{Linear}(\bm{\mathrm{h}}_{u-1}^{\mathsf{LE}}))\label{eq:3},
\\
&P(y_{u}|y_{1:u-1}, \bm{\mathrm{x}}_{1:t}) = \mathrm{SoftMax}(\bm{\mathrm{h}})\label{eq:4},
\end{align}
where the acoustic feature $\bm{\mathrm{h}}_{t}^{\mathsf{AE}}$ extracted from $\bm{\mathrm{x}}_{1:t}$ (Eq.~\eqref{eq:1})
and the feature $\bm{\mathrm{h}}_{u-1}^{\mathsf{LE}}$ from the previous output token sequence $y_{1:u-1}$ (Eq.~\eqref{eq:2}) 
are sent to the joint network, projected to the same dimension, and added up (Eq.~\eqref{eq:3}) to calculate the output probabilities against tokens in $\mathcal{V}$ based on the previous result (Eq.~\eqref{eq:4}).
Since the current symbol for each input frame is predicted based only on the past output tokens, streaming decoding is naturally introduced into the Transducer framework without additional effort.

In CBS-T, we utilize the CBS encoder as the acoustic encoder of a Transducer model to conduct streaming feature extraction along with Transducer streaming decoding, which achieves significant computational complexity reduction compared to CBS-E/D.

%% file: subtex/proposal.tex
\section{Proposal}
\label{sec:proposal}
To achieve high accuracy with zero look-ahead, we propose a multiple latency streaming ASR system, which leverages both a primary encoder with high latency to generate accurate outputs
and an auxiliary encoder to recognize the look-ahead frames attended by the primary encoder with no additional look-ahead.
In this section, we first describe the proposed system and then provide different methods for architecting the multiple latency streaming ASR.

\subsection{Multiple latency streaming ASR with CBS models}
Our proposed system can be constructed with both CBS-E/D and CBS-T streaming ASR models.
For block settings of the CBS encoder, we fix the size of history frames $N_l$ and target frames $N_c$ as $8$ and $4$, while the look-ahead range is controlled by choice of the look-ahead frame number $N_r$.
For instance, a primary encoder with the block setting of \textbf{8-4-4} ($N_l=8$, $N_c=4$, and $N_r=8$)
attends to four look-ahead frames, which induces a $128$ ms delay with a frame rate of $32$ ms.
Similarly, an auxiliary encoder with the block setting of \textbf{8-4-0} attends to a zero look-ahead frame.

Algorithm~\ref{proposed method} demonstrate how the proposed system works,
where the look-ahead frame number in the primary encoder equals the target frame number in the auxiliary encoder (i.e., $N_r=N'_c$).
During recognition, the primary encoder recognizes the target frames of the input block and outputs sequence $\textbf{y}^{p}_{c}$.  
Simultaneously, the auxiliary encoder recognizes the look-ahead frames of the input block and yields $\textbf{y}^{a}_{r}$.
$\textbf{y}^{p}_{c}$ is then appended to the previous output sequence $\textbf{y}^{P}$. 
At the appropriate timing, the auxiliary encoder outputs a special token $\langle /s \rangle$,
and the speech recognition process is terminated.
We concatenate the previous output sequence $\textbf{y}^{P}$ with the auxiliary encoder output $\textbf{y}^{a}_{r}$ as the final result.
Otherwise, the streaming ASR moves on to the next input block.

\begin{figure}[tb]
\centering
\centerline{\includegraphics[width=\linewidth]{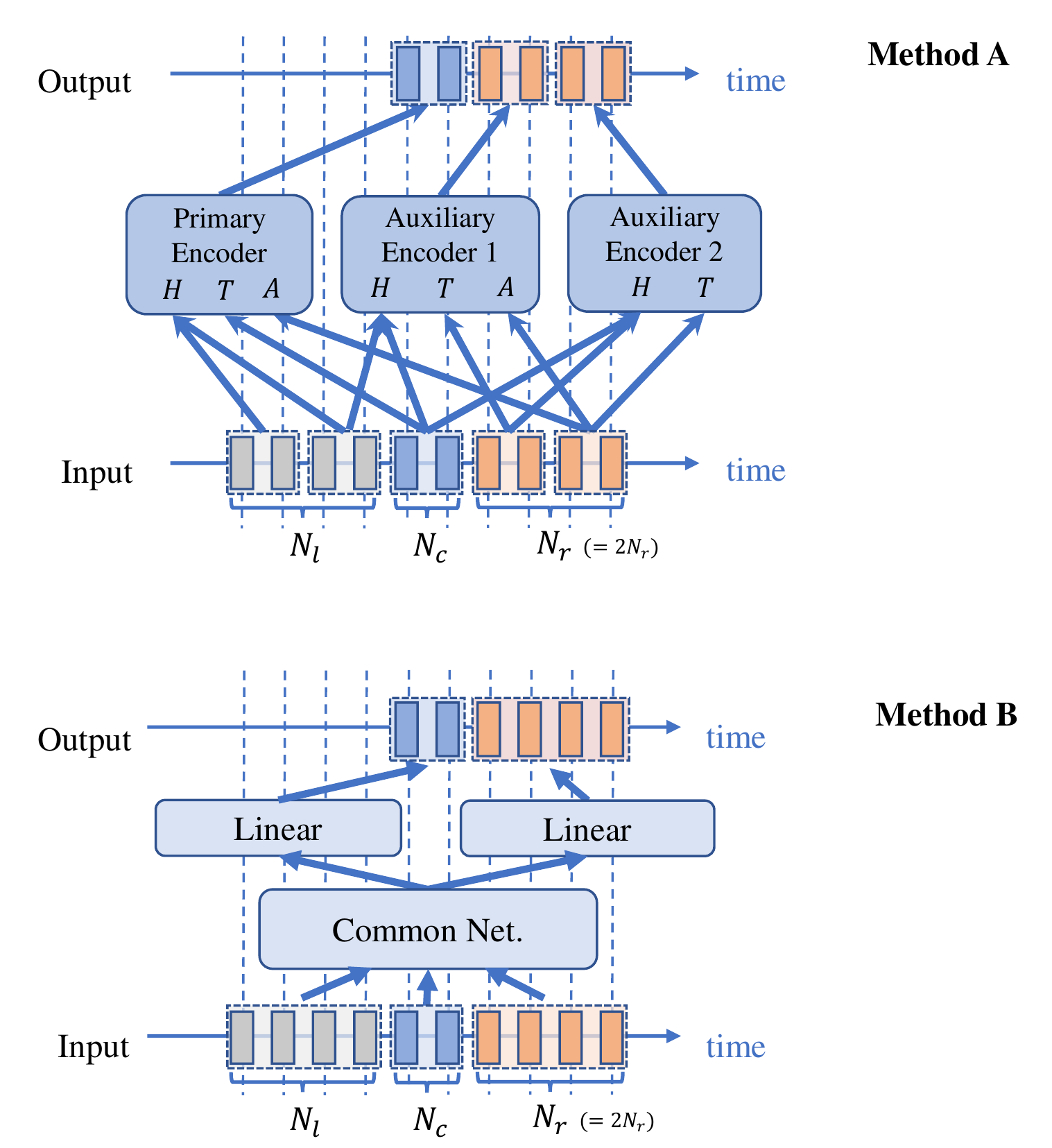}}
\caption{Structure of proposed system (case of $N_r=2 \times N_c$).
Data inputs of History, Target, and Look-ahead frames are indicated by arrows pointing at symbols H, T, and A, respectively.
In method A, primary encoder and auxiliary encoders share same structure.
}
\label{fig:proposal}
\end{figure}

Since the target frame of the auxiliary encoder $N'_c$ is fixed as four, the assumption of $N_r=N'_c$ constraints the look-ahead range of the primary encoder and limits the accuracy of the streaming ASR.
Therefore, we extend the two-encoder system in Algorithm~\ref{proposed method} with multiple auxiliary encoders, as shown in the upper part of Fig.~\ref{fig:proposal}.
Here the primary encoder attends to eight look-ahead frames,
which are recognized by two auxiliary encoders with four and zero look-ahead frames, respectively. 
With a more extensive look-ahead range ($256$ ms),
the streaming ASR achieves higher recognition accuracy while maintaining the operation with zero look-ahead.

\begin{algorithm}
\caption{Multi-latency streaming ASR}
\label{proposed method}
\begin{algorithmic}[1]
\State $T^{p}_{B} = N_l + N_c + N_r$      \Comment{primary encoder block setting}
\State $T^{a}_{B} = N'_l + N'_c$      \Comment{auxiliary encoder block setting}
\State $\textbf{y}^{P} \gets \emptyset$
\For{$t$ = $T^{p}_{B}$ to $T$ by $T^{p}_{B}$}
\State $\textbf{y}^{p}_{c}$ = PrimaryEncoder($X[t-T^{p}_{B}, t]$)
\State $\textbf{y}^{a}_{r}$ = AuxiliaryEncoder($X[t-T^{a}_{B}, t]$)
\State $\textbf{y}^{P} \gets \textbf{y}^{p}_{c}$ \Comment{extend target frame results} 
\If{$\langle /s \rangle$ in $\textbf{y}^{a}_{r}$}  \Comment{appropriate timing for ending}
\State break
\EndIf
\EndFor
\State $\textbf{y} = \textbf{y}^{P} +  \textbf{y}^{a}_{r} $      \Comment{final result}
\end{algorithmic}
\end{algorithm}

\subsection{Implementation methods}
The proposed multiple latency streaming ASR contains a primary encoder with look-ahead frames and an auxiliary encoder recognizing the look-ahead frames that operate in parallel.
The primary encoder is implemented following the same structure as the encoder in existing CBS models.
The auxiliary encoder shares parameters with the primary encoder but operates without attending to any look-ahead frames.
In this study, we propose two different methods to architect the proposed system, which are illustrated in Fig.~\ref{fig:proposal}.

\textbf{Method A (Parallel model).}
In Method A, the system contains $N_r / N_c$ auxiliary encoders in parallel, each of which shares exactly the same structure and parameters as the primary encoder.
The $i$-th auxiliary-encoder uses the input of the primary encoder shifted forward by $i \times N_c$ frames.
This means the look-ahead frames for the $i$-th encoder are shortened by $i \times N_c$ frames.
In the training phase, the primary encoder and all the auxiliary encoders are simultaneously trained by masking the last $N_c$, $2 \times N_c$, $\cdots$, $(N_r / N_c) \times N_c$ look-ahead frames with a certain probability.
This is expected to correctly encode the target part while enabling recognition of the look-ahead part without delay.

\textbf{Method B (Unified model).}
In Method B, we utilize a single network to recognize both target frames and look-ahead frames in one encoding pass.
With the block setting of $N_l$-$N_c$-$N_r$ , the model is trained to recognize $N_c$ only as well as recognize $N_c$ and $N_r$ together.
During inference, the model recognizes both $N_c$ and $N_r$ in one encoding pass,
where the results for $N_c$ are regarded as the primary encoder outputs, and the results for $N_r$ are used as the auxiliary encoder outputs.
Method B significantly reduces the computational cost during inference.
On the other hand, it solves more challenging problems than method A due to a large number of outputs.
The difficulty increases as the length of $N_r$ is extended.

%% file: subtex/exp.tex
\section{Experiments}
\label{sec:exp}
Speech recognition experiments were conducted on the proposed multiple latency streaming ASR system using ESPnet2~\cite{Watanabe2018ESPnetES, Boyer2021ASO}.

\subsection{Experimental setup}
The models were trained and evaluated using the Wall Street Journal (WSJ)~\cite{paul1992design} dataset.
We applied SpecAugment~\cite{Park2019SpecAugmentAS} to the input data for robust model training.
For the output tokens,
we used SentencePiece~\cite{kudo2018sentencepiece} to construct an 80 subword vocabulary from the training set with one additional token $\langle /s \rangle$.

We conducted experiments with both CBS-E/D and CBS-T models.
The CBS-E/D model consisted of a CBS encoder with six Conformer~\cite{Gulati2020ConformerCT} layers and decoder with six Transformer~\cite{vaswani2017attention} layers.
For CBS-T, we used a CBS encoder with six Conformer layers for the acoustic encoder and one layer of long short-term memory (LSTM) network~\cite{Hochreiter1997LongSM} for the label encoder.

All the models were trained by 150 epochs, and the final models were obtained by averaging the snapshots of the ten epochs with the best accuracy for CBS-E/D and minimal losses for CBS-T.
For decoding, a beam search was conducted with a beam size of ten for all.
We used the averaged word error rates (WER) on standard validation and test sets (\textit{dev93} and \textit{eval92}) to measure the recognition accuracy.

\begin{table}[tb]
\begin{center}
\caption{Word error rates on WSJ dataset.}
\label{table}
\resizebox{\columnwidth}{!}{
\small
\begin{tabular}{c c c c c}
    \toprule
     &  &   & \textbf{Delay} & \textbf{WER}\\
    \textbf{Model} & \textbf{Mode} & \textbf{Block Setting} & [ms] & [\%] ($\downarrow$) \\
    \midrule
    \multirow{7}{*}[-3pt]{CBS-E/D} & Single & 8-4-0 & 0 &  15.8  \\
    \cmidrule(l{0.5em}r{0.5em}){2-5}
    & Single & 8-4-4 & 128 & 14.3  \\
    &  Multiple (A) & 8-4-4, 8-4-0 & 0 & 14.8   \\
    &  Multiple (B) & 8-4-4, 12-4-0  & 0 & 14.8  \\
    \cmidrule(l{0.5em}r{0.5em}){2-5}
    & Single & 8-4-8  & 256 & 13.9  \\
    &  Multiple (A) & 8-4-8, 8-4-4, 8-4-0 & 0 & 14.3  \\
    &  Multiple (B) & 8-4-8, 12-8-0 & 0 & 14.2  \\
    \midrule
    \multirow{7}{*}[-3pt]{CBS-T}& Single & 8-4-0 & 0 & 16.4   \\
    \cmidrule(l{0.5em}r{0.5em}){2-5}
    & Single & 8-4-4 & 128 & 14.4  \\
    &  Multiple (A) & 8-4-4, 8-4-0 & 0 & 14.5  \\
    &  Multiple (B) & 8-4-4, 12-4-0 & 0 & 14.6  \\
    \cmidrule(l{0.5em}r{0.5em}){2-5}
     & Single & 8-4-8 & 256  & 13.7  \\
    &  Multiple (A) & 8-4-8, 8-4-4, 8-4-0 & 0 & \textbf{13.7} \\
    &  Multiple (B) & 8-4-8, 12-8-0 & 0 & 14.1  \\
    \bottomrule
\end{tabular}
}
\end{center}
\end{table}

\subsection{Experimental results}
The experimental results are summarized in Table~\ref{table}, 
where block settings are shown in the format of $N_l$-$N_c$-$N_r$, and the processing delay induced by look-ahead frames is recorded in the column of \textit{Delay}.
Results of the proposals are listed under mode names \textit{Multiple (A)} (Parallel model) and \textit{Multiple (B)} (Unified model). 
Baseline models are represented by mode name \textit{Single},
serving as upper-bounds (w/ look-ahead) and lower-bounds (w/ look-ahead).

Comparing the results of CBS-E/D and CBS-T, we can see that CBS-T showed inferior lower-bound results but outperformed CBS-E/D when look-ahead frames were applied.
Considering reducing decoding time, CBS-T showed higher suitability for our proposal.
Extending the primary encoder block setting from $8$-$4$-$4$ to $8$-$4$-$8$ vastly improved the performance of the streaming ASR.
With the setting of $8$-$4$-$8$, CBS-T achieved a WER of $13.7\%$,
which is very close to the non-streaming result with the same model structure ($12.8\%$, not shown in the table).
Regarding the different architectures of the proposal,
methods A and B presented the same level of performance in most cases, while for CBS-T, method A showed better accuracy with an extensive look-ahead range applied.
Hence, for implementing the proposed system, method A is more suitable when accuracy is more critical, while method B should be adopted when the computational cost comes first.
Overall, with the proposed multiple latency system, we managed to maintain high recognition accuracy while operating with zero look-ahead. 
Compared to the upper-bound result, zero performance degradation was achieved when applying method A to CBS-T with eight look-ahead frames.

%% file: subtex/conclusion.tex
\section{Conclusion}
\label{sec:conclusion}
In this study, we proposed a multiple latency streaming ASR system based on the CBS models to operate encoders with various latency in parallel.
Various implementation methods were studied for constructing the system.
Experimental results have shown our proposal's effectiveness in maintaining high recognition accuracy with zero look-ahead.
Our future work will focus on the trade-off between computational cost and recognition accuracy. We are also planning to incorporate the proposed model into the EoU (end-of-utterance)-detection-free turn-taking model~\cite{Sakuma2022} to realize a rhythmic conversation system.